\documentclass[a4,12pt,reqno]{article}
\usepackage[english]{babel}
\usepackage[latin1]{inputenc}
\usepackage{graphics}
\usepackage{ulem}
\usepackage{hhline}
\usepackage{dsfont}
\usepackage{mathrsfs}
\usepackage{fancyhdr}
\usepackage{amsmath,amssymb}
\usepackage{rotating}
\usepackage[latin1]{inputenc}
\usepackage[T1]{fontenc}
\usepackage{fancybox}
\usepackage{color}
\usepackage{colortbl}
\usepackage{setspace}
\usepackage{enumerate}
\usepackage[nice]{nicefrac}
\usepackage{amsthm}
\usepackage{multicol}
\usepackage{pifont}
\usepackage[french]{minitoc}
\usepackage{latexsym,amsfonts}
\usepackage{amsthm}
\usepackage{varioref}
\usepackage{textcomp}
\usepackage{lmodern}
\usepackage{mathpazo}
\usepackage{euscript}
\usepackage[pdftex]{hyperref}
\numberwithin{equation}{section}
\makeatletter\@addtoreset{equation}{section}

\DeclareMathSymbol{\subsetneqq}{\mathbin}{AMSb}{36}

\begin{document}

\author{Zouha\"{i}r MOUAYN}
\title{\textsf{A generating function for Hermite polynomials in
connection with Euclidean Landau levels}}
\date{{\small Department of Mathematics, Faculty of Sciences and Technics
(M'Ghila), Sultan Moulay Slimane University, BP 523, B\'{e}ni Mellal,
Morocco }\\
{\small (e-mail: mouayn@gmail.com)}}
\maketitle

\begin{abstract}
We have formulated a generating function for the Hermite polynomials by
comparing two expressions of the same coherent states attached to planar
Landau levels. A first expression is obtained by generalizing the canonical
coherent states when written as series expansion in the basis of number
states. While the second expression is established by following a
construction based on group representation theory.
\end{abstract}

\section{Introduction}

An important class of orthogonal polynomial encountered in the applications,
especially in mathematical physics, consists of the Hermite polynomials $%
H_{n}\left( \xi \right) $ which can be defined by the Rodriguez formula as
\begin{equation}
H_{n}\left( \xi \right) =\left( -1\right) ^{n}e^{\xi ^{2}}\frac{d^{n}}{d\xi
^{n}}\left( e^{-\xi ^{2}}\right) , \quad n=0,1,2,\cdots .   \label{1.1}
\end{equation}
According to \eqref{1.1}, the first few Hermite polynomials are
\begin{equation}
H_{0}\left( \xi \right) =1, \quad H_{1}\left( \xi \right) =2\xi , \quad H_{2}\left( \xi
\right) =4\xi ^{2}-2,\cdots ,  \label{1.2}
\end{equation}
and in general,
\begin{equation}
H_{n}\left( \xi \right) =\sum_{k=0}^{\left[ n/2\right] }\frac{\left(
-1\right) ^{k}n!}{k!\left( n-2k\right) !}\left( 2\xi \right) ^{n-2k},
\label{1.3}
\end{equation}
where $[ \eta ] $ denotes the largest integer $\leq \eta .$ These
polynomials multiplied by the constant factor $1/n!$ are the coefficients in
the expansion
\begin{equation}
\exp \left( 2\xi \tau -\tau^{2}\right) =\sum_{n=0}^{+\infty }\frac{H_{n}\left(
\xi \right) }{n!}\tau ^{n},  \label{1.4}
\end{equation}
called the generating function of the Hermite polynomials \cite{1}.

Here, our aim is to draw attention to an interesting relation satisfied by
the Hermite polynomials, which arise when comparing two expressions of the
same coherent states attached to planar Landau levels. The relation we
propose here is a kind of generalization of \eqref{1.4} involving
the polynomials
\begin{equation}
L_{m}^{(k)}\left( x\right) :=\sum\limits_{j=\max \left( 0,-k\right) }\left(
-1\right) ^{j}\frac{\left( k+m\right) !}{j!\left( m-k\right) !\left(
k+j\right) !}x^{j}  \label{1.5}
\end{equation}
which correspond to the usual Laguerre polynomial for $k\geq 0$. However,
equation \eqref{1.5} is valid for negative value of $k$ \cite[p. 98]{16}. Precisely, we establish the following formula:

\noindent\textbf{Theorem 1.1. }\textit{Let }$m=0,1,2,\cdots ,\beta >0$\textit{\ and }$%
a,b\in \Bbb{R}$. \textit{Then},
\begin{align}
\sum_{k=-m}^{+\infty }\frac{\left( \sqrt{\beta }\left( a+ib\right) \right)
^{k}}{2^{k}\left( k+m\right) !} &L_{m}^{(k)}\left( \frac{\beta }{2}\left(
a^{2}+b^{2}\right) \right) H_{k+m}\left( \xi \right) \nonumber \\&
=\frac{1}{m!}\exp \left( -\frac{\beta }{4}\left( a-ib\right) ^{2}+\xi \left(
a-ib\right) \sqrt{\beta }\right) H_{m}\left( \xi -a\sqrt{\beta }\right).
\label{1.6}
\end{align}
Note that for $m=0,$ $b=0$ and $a= 2\tau \beta^{-1/2},$ the
relation \eqref{1.6} reduces to \eqref{1.4}. I didn't find  formula \eqref{1.6} in books dealing with special functions and orthogonal polynomials but after sending this manuscript to many colleagues I have received in response a letter \cite{Ismail} from Professor Mourad ISMAIL who has established a direct proof for this formula \eqref{1.6}. I would like to thank him very much for reading the manuscript and for his interest on this work.

This paper is organized as follows. In Section 2, we recall some facts on
the canonical coherent states of the quantum harmonic oscillator as well as
two types of their generalization we will be dealing with. In Section 3, we
summarize some needed tools on the planar Landau levels. In Section 4,  we
construct coherent states via a group representation theory. In Section 5,
we rederive these coherent states via a series expansion.

\section{The canonical coherent states}

The first model of coherent states was the `nonspreading wavepacket' of the
harmonic oscillator, which have been constructed by Schr\"{o}dinger \cite{2}. In suitable units, wave functions of these states can be written
as
\begin{equation}
\Phi _{\frak{z}}\left( \xi \right) :=\left\langle \xi \mid \frak{z}%
\right\rangle =\pi ^{-\frac{1}{2}}\exp \left( -\frac{1}{2}\xi ^{2}+\sqrt{2}%
\xi \frak{z}-\frac{1}{2}\frak{z}^{2}-\frac{1}{2}\left| \frak{z}\right|
^{2}\right) ,\xi \in \Bbb{R},  \label{2.1}
\end{equation}
where $\frak{z}\in \Bbb{C}$ determines the mean values of coordinate $%
\widehat{x}$ and momentum $\widehat{p}$ according to $\left\langle \widehat{x%
}\right\rangle :=\left\langle \Phi _{\frak{z}},x\Phi _{\frak{z}%
}\right\rangle =\sqrt{2}{\Re}\frak{z}$ and $\left\langle \widehat{p}%
\right\rangle :=\left\langle \Phi _{\frak{z}},p\Phi _{\frak{z}}\right\rangle
=\sqrt{2}{\Im}\frak{z}.$ The variances $\sigma _{x}=\left\langle
\widehat{x}^{2}\right\rangle -\left\langle \widehat{x}\right\rangle ^{2}=%
\frac{1}{2}$ and $\sigma _{p}=\left\langle \widehat{p}^{2}\right\rangle
-\left\langle \widehat{p}\right\rangle ^{2}=\frac{1}{2}$ have equal values,
so their product assumes the minimal value permitted by the Heisenberg
uncertainty relation.

\ The coherent state $\Phi _{\frak{z}}$ have been also obtained by Feymann
\cite{3}   and Glauber \cite{4}  from the vacuum state $\mid
0\rangle $ by means of the unitary displacement operator $\exp \left( \frak{z%
}A^{\ast }-\overline{\frak{z}}A\right) $ as
\begin{equation}
\Phi _{\frak{z}}=\exp \left( \frak{z}A^{\ast }-\overline{\frak{z}}A\right)
\mid 0\rangle ,  \label{2.2}
\end{equation}
where $A$ and $A^{\ast }$ are annihilation and creation operators defined by
\begin{equation}
A=\frac{1}{\sqrt{2}}\left( \widehat{x}+i\widehat{p}\right) , \qquad  A^{\ast
}=\frac{1}{\sqrt{2}}\left( \widehat{x}-i\widehat{p}\right) .  \label{2.3}
\end{equation}
It was Iwata \cite{5}  who used the well known expansion over the
Fock basis $\mid n\rangle $ to give an expression of $\Phi _{\frak{z}}$ as
\begin{equation}
\Phi _{\frak{z}}=e^{-\frac{1}{2}\left| \frak{z}\right|
^{2}}\sum\limits_{n=0}^{+\infty }\frac{\frak{z}^{n}}{\sqrt{n!}}\mid n\rangle
.  \label{2.4}
\end{equation}
Actually, various generalizations of coherent states have been proposed.
Here, we shall focus on two of them: $(i)$ a group-theoretic
generalization \textit{''\`{a} la Perelomov''}of \eqref{2.2} in
which coherent states are produced by the action $T_{g}$ of the group
element $g\in G$ on a reference state $\phi _{0}$ in a representation
Hilbert space $\mathcal{H}$ as (\cite{6}):
\begin{equation}
\widetilde{\Phi }_{\frak{g}}=T_{g}\left[ \phi _{0}\right] ,  \label{2.5}
\end{equation}
$\left( i\right) $ a generalization \textit{''\`{a} la Iwata'' of } \eqref{2.4}, by choosing different set of coefficients $%
\left\{ c_{n}\left( \frak{z}\right) \right\} $ and functions $\left\{
\varphi _{n}\right\} $ satisfying suitable conditions as
\begin{equation}
\widetilde{\Phi }_{\frak{z}}=\sum\limits_{n}c_{n}\left( \frak{z}\right) \mid
\varphi _{n}\rangle .  \label{2.6}
\end{equation}
For an overview of all aspect of the theory of coherent states we refer to
the survey of V.V. Dodonov \cite{7}.

\section{ Landau levels on $\Bbb{R}^{2}$}

The Hamiltonian of a single electron, moving on a two-dimensional plane
subject to a uniform magnetic field along the $z$-direction, is given by
\begin{equation}
H_{\beta }:=\frac{1}{2}\left( i\partial _{x}-\frac{\beta }{2}y\right) ^{2}+%
\frac{1}{2}\left( i\partial _{y}+\frac{\beta }{2}x\right) ^{2},  \label{3.1}
\end{equation}
where $\beta >0$ is proportional to the field strength. $H_{\beta }$\ is an
unbounded symmetric operator on $C_{0}^{\infty }(\Bbb{R}^{2})$ and\ is an
essentially self-adjoint operator in $L^{2}(\Bbb{R}^{2},d\mu \left(
x,y\right) )$, $d\mu $ being the Lebesgue measure on $\Bbb{R}^{2}.$  The
spectrum\ of $H_{\beta }$ consists of eigenvalues of infinite multiplicity (%
\textit{Landau levels}) of the form
\begin{equation}
\epsilon _{m}^{\beta }:=(m+\frac{1}{2})\beta , \quad m=0,1,2, \cdots .  \label{3.2}
\end{equation}
We let $\mathcal{A}_{m,\beta }\left( \Bbb{R}^{2}\right) $ denote the
eigensubspace of the operator $H_{\beta }$, which corresponds to the
eigenvalue $\epsilon _{m}^{\beta }$ in \eqref{3.2} as
\begin{equation}
\mathcal{A}_{m,\beta }\left( \Bbb{R}^{2}\right) :=\left\{ \varphi \in L^{2}(\Bbb{R}%
^{2}),H_{\beta }\varphi =\epsilon _{m}^{B}\varphi \right\} .  \label{3.3}
\end{equation}
Our next goal is to consider a basis of the subspace in \eqref{3.3}.
For $\mathbf{(}x,y)\in \Bbb{R}^{2},$ and $k=-m,-m+1,\cdots ,$ set
\begin{equation}
\varphi _{k}^{\beta ,m}\mathbf{(}x,y):=c_{\beta ,m,k}(\sqrt{\frac{\beta }{2}}%
(x+iy))^{k}e^{-\frac{\beta }{4}(x^{2}+y^{2})}L_{m}^{(k)}(\frac{\beta }{2}%
(x^{2}+y^{2})),  \label{3.4}
\end{equation}
where
\begin{equation}
c_{\beta ,m,k}=\left( \frac{m!\beta }{2\pi (k+m)!}\right) ^{\frac{1}{2}}.
\label{3.5}
\end{equation}
The functions $\varphi _{k}^{\beta ,m},k=-m,-m+1,\cdots ,$ constitute an
orthonormal basis of the Hilbert space $\mathcal{A}_{m,\beta }\left( \Bbb{R}%
^{2}\right) $ whose reproducing kernel is given by
\begin{equation}
K_{m,\beta }(\left( x,y\right) ,\left( x^{\prime },y^{\prime }\right)
):=\sum_{k=-m}^{+\infty }\varphi _{k}^{\beta ,m}\left( x,y\right) \overline{%
\varphi _{k}^{\beta ,m}\left( x^{\prime },y^{\prime }\right) }.  \label{3.6}
\end{equation}
Explicitly, we have that (\cite[p. 1058]{8}):
\begin{equation}
K_{m,\beta }(\mathbf{r},\mathbf{r}^{\prime })=\frac{\beta }{2\pi }\exp
\left( -\frac{i\beta }{2}\mathsf{\ }\mathbf{r}\wedge \mathbf{r}^{\prime }-%
\frac{\beta }{4}\left\| \mathbf{r}-\mathbf{r}^{\prime }\right\| ^{2}\right)
L_{m}^{(0)}\left( \frac{\beta }{2}\left\| \mathbf{r}-\mathbf{r}^{\prime
}\right\| ^{2}\right) .  \label{3.7}
\end{equation}
where $\mathbf{r=(}x,y),\mathbf{r}^{^{\prime }}\mathbf{=(}x^{^{\prime
}},y^{^{\prime }})\in \Bbb{R}^{2}$ and $\mathbf{r}\wedge \mathbf{r}^{\prime
}=xy^{^{\prime }}-x^{\prime }y.$

For more information on Landau levels we refer the reader to \cite{9}.

\section{Perelomov's coherent states}

As pointed out in Section 2, various generalizations of the notion of
coherent states have been proposed. Here, for our purpose, we adopt the so
called Prelomov's approach based on a group-theoretic formalism  \cite{6}.

For this, let $\Bbb{H}_{1}$ be the Heisenberg group (of degree $1$) i.e. the
Lie group whose underlying manifold is $\Bbb{R}^{3}$ with coordinates ($%
x,y,t)$\ and whose group law is
\begin{equation}
(x,y,t).(x^{\prime },y^{\prime },s)=(x+x^{\prime },y+y^{\prime },t+s+\frac{1%
}{2}(xy^{^{\prime }}-x^{\prime }y)).  \label{4.1}
\end{equation}
It is standard that Lebesgue measure on $\Bbb{R}^{3}$\ is bi-invariant Haar
measure on $\Bbb{H}_{1}.$ The continuous unitary irreducible representations
of $\Bbb{H}_{1}$ are well known \cite[p. 37]{10}. Here, we will
be concerned with a representation of $\Bbb{H}_{1}$ on the Hilbert space $%
L^{2}(\Bbb{R},d\xi )$ by shift and multiplication operators \cite[\S 1.1]{11}:
\begin{equation}
T_{\beta }(x,y,t)\left[ \psi \right] (\xi ):=\exp i\left( \beta t-\sqrt{%
\beta }y\xi +\frac{\beta }{2}xy\right) \psi \left( \xi -\sqrt{\beta }%
x\right)  \label{4.2}
\end{equation}
for $(x,y,t)\in \Bbb{H}_{1}\mathbf{,}$ $\beta >0,$ $\psi \in L^{2}(\Bbb{R}%
,d\xi )$ and $\xi \in \Bbb{R}$, called the Schr\"{o}dinger representation.\

\noindent \textbf{Proposition 4.1.}\textit{The Schr\"{o}dinger representation }$T_{\beta }$\textit{\ in } \eqref{4.2} \textit{satisfies the equality:}
\begin{equation}
\int\limits_{\Bbb{R}^{2}}\left\langle \psi _{1},T_{\beta }(x,y,0)\left[ \phi
_{1}\right] \right\rangle \left\langle T_{\beta }(x,y,0)\left[ \phi _{2}%
\right] ,\psi _{2}\right\rangle d\mu (x,y)=\left\langle \psi _{1},\psi
_{2}\right\rangle \left\langle \phi _{1},\phi _{2}\right\rangle   \label{4.3}
\end{equation}
\textit{for all }$\psi _{1},\psi _{2}\in L^{2}(\Bbb{R},d\xi )$\textit{\ and }%
$\phi _{1},\phi _{2}\in L^{2}(\Bbb{R},d\xi ).$

\begin{proof} By \cite{12}, the unitary irreducible
representation $T_{\beta }$\ in \eqref{4.2} is square integrable modulo: the
center $\Bbb{R}$ of $\Bbb{H}_{1}$ and the Borelian section $\sigma _{0}$\
defined from the coset space $\Bbb{H}_{1}/\Bbb{R}\mathbf{=}\Bbb{R}^{2}$ into
$\Bbb{H}_{1}$ by $\sigma _{0}(x,y):=(x,y,0)$. Next, by theorem \cite{13} there exists operator $\delta $\ in $L^{2}(\Bbb{R},d\xi )$%
\ self-adjoint, positive and semi-invariant such that
\begin{equation}
\int\limits_{\Bbb{R}^{2}}\left\langle \psi _{1},T_{\beta }(x,y,0)\left[ \phi
_{1}\right] \right\rangle \left\langle T_{\beta }(x,y,0)\left[ \phi _{2}%
\right] ,\psi _{2}\right\rangle d\mu (x,y)=\left\langle \psi _{1},\psi
_{2}\right\rangle \left\langle \delta ^{\frac{1}{2}}\phi _{1},\delta ^{\frac{%
1}{2}}\phi _{2}\right\rangle   \label{4.4}
\end{equation}
for all $\psi _{1},\psi _{2}\in L^{2}(\Bbb{R},d\xi )$ and $\phi _{1},\phi
_{2}\in Dom(\delta ^{\frac{1}{2}}).$ Here, the group $\Bbb{H}_{1}$ is
unimodular therefore the operator $\delta $\ is the identity (\cite[p. 215]{13}).
\end{proof}

\noindent  \textbf{Definition 4.1}. \textit{Let} $m=0,1,\cdots ,$\textit{\ and }$\beta >0.$
\textit{The Perelomov's coherent states attached to the Landau level }$%
\epsilon _{m}^{\beta }$ \textit{in \eqref{3.2} and labeled by points (}$x,y)$%
\textit{\ of the coset space }$\Bbb{R}^{2}$\textit{\ are defined by }
\begin{equation}
\Phi _{(x,y),\beta ,m}:=T_{\beta }(\sigma _{0}(x,y))\left[ \phi _{m}\right] .
\label{4.5}
\end{equation}
\textit{where the reference state }$\phi _{m}$ \textit{\ is given by the
Gaussian-Hermite function}
\begin{equation}
\phi _{m}(\xi ):=\left( \sqrt{\pi }2^{m}m!\right) ^{-\frac{1}{2}}e^{-\frac{1%
}{2}\xi ^{2}}H_{m}(\xi ),\xi \in \Bbb{R}.  \label{4.6}
\end{equation}
\textit{Their wave functions are given by}
\begin{equation}
\Phi _{(x,y),\beta ,m}(\xi )=\left( \sqrt{\pi }2^{m}m!\right) ^{-\frac{1}{2}%
}e^{-i\sqrt{\beta }\xi y+i\frac{\beta }{2}xy-\frac{1}{2}\left( \xi -\sqrt{%
\beta }x\right) ^{2}}H_{m}\left( \xi -\sqrt{\beta }x\right) .  \label{4.7}
\end{equation}
\textbf{Proposition 4.2.}\textit{\ The unity of the Hilbert space }$L^{2}(%
\Bbb{R},d\xi )$\textit{\ is solved a}s
\begin{equation}
\mathbf{1}_{\mathsf{\ }L^{2}(\Bbb{R},d\xi )}=\int\limits_{\Bbb{R}^{2}}d\mu
(x,y)\mid (x,y),\beta ,m><m,\beta ,(x,y)\mid   \label{4.8}
\end{equation}
\textit{where the ket vector}$\mid (x,y),\beta ,m>\equiv \Phi _{(x,y),\beta
,m}$.

\begin{proof} The coherent states in \eqref{4.5}  are completely
justified by the square integrability of the unitary irreducible
representations $T_{\beta }$\ modulo the subgroup $\Bbb{R}$ and the section $%
\sigma _{0}$.\ In this context Eq. \eqref{4.3} reads
\begin{equation}
\int\limits_{\Bbb{R}^{2}}<\psi \mathsf{,}\Phi _{(x,y),\beta ,m}><\Phi
_{(x,y),\beta ,m},\psi >d\mu (x,y)=\left\langle \psi ,\psi \right\rangle
,\quad \psi \in L^{2}(\Bbb{R},d\xi ).  \label{4.9}
\end{equation}
The square integrability of the representation $T_{\beta }$ \textit{mod}($%
\sigma _{0},\Bbb{R}$) can be reformulated as a resolution of the identity.
Indeed, denoting $\Phi _{(x,y),\beta ,m}=\mid (x,y),B,m>,$ one can use the
Dirac's\textit{\ bra-ket} notation to write \eqref{4.4} in the form
\eqref{4.8}. \end{proof}

\section{Iwata's coherent states}

In this section, we present a generalization of \ the coherent states as
given by Iwata's series \eqref{2.4} according to the procedure in
\cite{14}.

Let $(X,\nu )$\ be a measure space and let $\mathcal{A}\subset L^{2}(X,\nu )$%
\ be a closed subspace of infinite dimension. Let $\left\{ f_{n}\right\}
_{n=1}^{\infty }$ be an orthogonal basis of $\mathcal{A}$ satisfying, for
arbitrary $u\in X,$
\begin{equation}
\omega \left( u\right) :=\sum_{n=1}^{\infty }\rho _{n}^{-1}\left|
f_{n}\left( u\right) \right| ^{2}<+\infty ,  \label{5.1}
\end{equation}
where $\rho _{n}:=\left\| f_{n}\right\| _{L^{2}(X)}^{2}$. Define
\begin{equation}
K(u,v):=\sum_{n=1}^{\infty }\rho _{n}^{-1}f_{n}(u)\overline{f_{n}(v)},\quad
u,v\in X.  \label{5.2}
\end{equation}
Then, $K(u,v)$\ is a reproducing kernel, $\mathcal{A}$ is the corresponding
reproducing kernel Hilbert space and $\omega \left( u\right) =K(u,u)$, $u\in
X$.\\

\noindent \textbf{Definition 5.1. } \textit{Let }$\mathcal{H}$\textit{\ be a Hilbert
space with }$\dim \mathcal{H}=\infty $\textit{\ and }$\left\{ \phi
_{n}\right\} _{n=1}^{\infty }$\textit{\ be an orthonormal basis of }$%
\mathcal{H}.$\textit{\ Therefore, for }$x\in X,$\textit{\ define}
\begin{equation}
\mid u>:=(\omega (u))^{-\frac{1}{2}}\sum_{n=1}^{\infty }\frac{f_{n}(u)}{%
\sqrt{\rho _{n}}}\phi _{n}.\quad \quad  \label{5.6}
\end{equation}
\textit{The vectors }$\left( \mid u>\right) _{u\in X}$\textit{\ will be
called coherent states of Iwata type.}\\

The choice of the Hilbert space $\mathcal{H}$\ defines a quantization of the
set $X=$\ $\left\{ u\right\} $ by the coherent states $\mid u>,$ via the
inclusion map $:u\rightarrow \mid u>$\ from $X$\ into $\mathcal{H}$. By
definition, it is straightforward to show that $<u\mid u>=1.$ This also
leads to the following definition:\\

\noindent \textbf{Definition 5.2. }\textit{The isometric map} $W:\mathcal{H\rightarrow
A}\subset L^{2}(X,\nu )$ \textit{defined by }$W[\phi ](u):=(\omega (u))^{%
\frac{1}{2}}<u\mid \phi >$ \textit{is called the coherent state transform
associated with the set of coherent states }$\left( \mid u>\right) _{u\in X}.$\\

Thus, for $\phi ,\psi \in \mathcal{H}$, we have
\begin{equation}
<\phi \mid \psi >_{\mathcal{H}}=<W\left[ \phi \right] \mid W\left[ \psi %
\right] >_{L^{2}\left( X\right) }=\int\limits_{X}d\nu \left( u\right) \omega
\left( u\right) <\phi \mid u><u\mid \psi >  \label{5.7}
\end{equation}
and thereby we have a resolution of the identity
\begin{equation}
\mathbf{1}_{\mathcal{H}}=\int\limits_{X}d\nu \left( u\right) \omega \left(
u\right) \mid u><u\mid ,  \label{5.8}
\end{equation}
where $\omega \left( u\right) $\ appears as a weight function.\\

\noindent \textbf{Definition} \textbf{5.3.}\textit{For each Landau level }$\epsilon
_{m}^{\beta }$\textit{\ in \eqref{3.2} define a set of coherent states according
to formula \eqref{5.6} as }
\begin{equation}
\mid (x,y),\beta ,m>:=\left( K_{\beta ,m}\left( \left( x,y\right) ,\left(
x,y\right) \right) \right) ^{-\frac{1}{2}}\sum_{n=0}^{+\infty }\frac{\varphi
_{n}^{\beta ,m}(x,y) }{\sqrt{\rho _{n}^{\beta ,m}\ }}\phi _{n},
\label{5.9}
\end{equation}
\textit{where }$K_{\beta ,m}\left( .,.\right) $ \textit{is the reproducing
kernel given in \eqref{3.7}, }$\varphi _{n}^{\beta ,m}\left( x,y\right) $\textit{\
are the functions given in \eqref{3.4} and } $$\phi _{n}\left( \xi \right) =\left(
\sqrt{\pi }2^{n}n!\right) ^{-\frac{1}{2}}e^{-\frac{1}{2}\xi ^{2}}H_{n}(\xi ).$$

It is well known \cite{15} that the functions $\phi _{n}(\xi)$, $n=0,1,\cdots ,$ constitute an orthonormal basis of the
Hilbert space $L^{2}\left( \Bbb{R},d\xi \right) $. The latter carries the
constructed coherent states in \eqref{5.9}.\\

\noindent \textbf{Proposition 5.1. }\textit{The wave functions of the states }\eqref{5.9}
\textit{\ can be expressed in a closed form by the expression in } \eqref{4.7}  \textit{. i.e., the equality}
\[ <\xi \mid \left( x,y\right) ,\beta ,m>=\mathit{\ }\Phi _{(x,y),\beta ,m}(\xi).\]
holds for any $\xi \in \Bbb{R}.$

\begin{proof} According to Eq. \eqref{5.9}, we start by writing
\begin{align}
<\xi &\mid \left( x,y\right) ,\beta ,m> :=\left( \frac{\beta }{2\pi }\right)
^{-\frac{1}{2}}\sum_{n=0}^{+\infty }\varphi _{n}^{\beta ,m}\left( x,y\right)
\phi _{n}\left( \xi \right)  \label{5.11} \\
& =e^{-\frac{\beta }{4}\left( x^{2}+y^{2}\right) }\sum_{n=0}^{+\infty }\sqrt{%
\frac{m!}{n!}}\left( \sqrt{\frac{\beta }{2}}\right) ^{n-m}\left( x+iy\right)
^{n-m}L_{m}^{(n-m)}\left( \frac{\beta \left( x^{2}+y^{2}\right) }{2}\right)
\phi _{n}\left( \xi \right)   \label{5.12}
\end{align}
But since $n$ takes values from $0$ to $+\infty $, we have to take care when
$0\leq n<m.$ In this case we make use of  Eq. \eqref{1.5} or \cite[p. 98]{16} to write
\begin{equation}
L_{m}^{\left( -p\right) }\left( w\right) =\left( -w\right) ^{p}\frac{\left(
m-p\right) !}{m!}L_{m-l}^{\left( p\right) }\left( w\right) , \quad 1\leq p\leq m,
\label{5.13}
\end{equation}
for $p=m-n$ to write the Laguerre polynomial with upper index $n-m<0$ as
\begin{equation}
L_{m}^{(n-m)}\left( \frac{\beta \left( x^{2}+y^{2}\right) }{2}\right)
=\left( -\frac{\beta }{2}\left( x^{2}+y^{2}\right) \right) ^{m-n}\frac{n!}{m!%
}L_{n}^{\left( m-n\right) }\left( \frac{\beta \left( x^{2}+y^{2}\right) }{2}%
\right) .  \label{5.14}
\end{equation}
Next we make use of the identity \cite[p. 838]{1}
\begin{equation}
\int\limits_{\mathbf{-\infty }}^{+\infty }e^{-u^{2}}H_{s}(u+\alpha
)H_{l}(u+\gamma )du=2^{l}\sqrt{\pi }s!\gamma ^{l-s}L_{s}^{(l-s)}(-2\alpha
\gamma ),   \quad s\leq l , \label{5.15}
\end{equation}
for
\begin{equation}
-2\alpha \gamma =\frac{\beta }{2}\left( x^{2}+y^{2}\right) ,~\alpha =-\frac{%
\sqrt{\beta }}{2}\left( x-iy\right), \qquad \gamma =\frac{\sqrt{\beta }}{%
2}\left( x+iy\right).  \label{5.16}
\end{equation}
After calculations, the following identity holds for any\ $n=0,1,2, \cdots,$%
\begin{equation}
\left( x+iy\right) ^{n-m}L_{m}^{(n-m)}\left( \frac{\beta }{2}\left(
x^{2}+y^{2}\right) \right) =\frac{\left( \sqrt{\beta }\right) ^{m-n}}{2^{m}m!%
\sqrt{\pi }}\int\limits_{\mathbf{-\infty }}^{+\infty
}e^{-u^{2}}H_{m}(u+\alpha )H_{n}(u+\gamma )du . \label{5.17}
\end{equation}
Returning back to Eq. \eqref{5.12}, it follows that
\begin{align}
<\xi & \mid ( x,y) ,\beta ,m>  \nonumber \\
 & =e^{-\frac{\beta }{4}\left(
x^{2}+y^{2}\right) } \sum_{n=0}^{+\infty }\sqrt{\frac{m!}{n!}}\left( \frac{\left( \sqrt{2}%
\right) ^{n-m}}{2^{n}m!\sqrt{\pi }}\int\limits_{\mathbf{-\infty }}^{+\infty
}e^{-u^{2}}H_{m}(u+\alpha )H_{n}(u+\gamma )du\right) \phi _{n}\left( \xi
\right)   \label{5.19}\\
&=\frac{e^{-\frac{\beta }{4}\left( x^{2}+y^{2}\right) }}{\left( \sqrt{2}%
\right) ^{m}\sqrt{\pi m!}}\sum_{n=0}^{+\infty }\frac{1}{\sqrt{n!}\left(
\sqrt{2}\right) ^{n}}\left( \int\limits_{\mathbf{-\infty }}^{+\infty
}e^{-u^{2}}H_{m}(u+\alpha )H_{n}(u+\gamma )du\right) \phi _{n}\left( \xi
\right) .  \label{5.19}
\end{align}
So, we have to consider the integral
\begin{equation}
I_{m,n}=\int\limits_{\mathbf{-\infty }}^{+\infty }e^{-u^{2}}H_{m}(u+\alpha
)H_{n}(u+\gamma )du.  \label{5.20}
\end{equation}
By the change of variable: $u\rightarrow u-\gamma ,$ we obtain that
\begin{equation}
I_{m,n}=\int\limits_{\mathbf{-\infty }}^{+\infty }\exp \left(
-u^{2}+2u\gamma -\gamma ^{2}\right) H_{m}(u+\alpha -\gamma )H_{n}(u)du .
\label{5.21}
\end{equation}
But, since
\begin{equation}
\alpha -\gamma =-\frac{\sqrt{\beta }}{2}\left( x-iy\right) -\frac{\sqrt{%
\beta }}{2}\left( x+iy\right) =-\sqrt{\beta }x,  \label{5.22}
\end{equation}
we get successively
\begin{align}
I_{m,n} &=\int\limits_{\mathbf{-\infty }}^{+\infty }\exp \left(
-u^{2}+2u\gamma -\gamma ^{2}\right) H_{m}(u-\sqrt{\beta }x)H_{n}(u)du
\label{5.23}
\\&  =   \int\limits_{\mathbf{-\infty }}^{+\infty }e^{-u^{2}+2u\frac{\sqrt{\beta }}{2%
}\left( x+iy\right) -\left( \frac{\sqrt{\beta }}{2}\left( x+iy\right)
\right) ^{2}}H_{m}(u-\sqrt{\beta }x)H_{n}(u)du  \label{5.24}
\\&  =  \int\limits_{\mathbf{-\infty }}^{+\infty }e^{i\sqrt{\beta }uy-i\frac{\beta
}{2}xy-\frac{\beta }{4}x^{2}-u^{2}+u\sqrt{\beta }x}e^{\frac{\beta }{4}%
y^{2}}H_{m}(u-\sqrt{\beta }x)H_{n}(u)du.  \label{5.25}
\end{align}
So that
\begin{align}
<\xi \mid & \left( x,y\right) ,\beta ,m> \nonumber \\
=&\frac{\left( \sqrt{2}\right) ^{-m}}{\sqrt{m!\pi }}\sum_{n=0}^{+\infty }%
\frac{1}{\sqrt{n!}}\left( \frac{1}{\left( \sqrt{2}\right) ^{n}}\int\limits_{%
\mathbf{-\infty }}^{+\infty }e^{-u^{2}}H_{m}(u+\alpha )H_{n}(u+\gamma
)du\right) \phi _{n}\left( \xi \right)    \label{5.26}
\\
=&\frac{1}{\left( \sqrt{2}\right) ^{m}\sqrt{\pi }\sqrt{m!}}%
\sum_{n=0}^{+\infty }\frac{1}{\sqrt{n!}}\frac{\phi _{n}\left( \xi \right) }{%
\left( \sqrt{2}\right) ^{n}}  \label{5.27} \\
&\qquad \qquad  \quad \times \int\limits_{\mathbf{-\infty }}^{+\infty }e^{-\frac{1}{2}u^{2}}e^{i%
\sqrt{\beta }uy-i\frac{\beta }{2}xy-\frac{\beta }{2}x^{2}-\frac{1}{2}u^{2}+u%
\sqrt{\beta }x}H_{m}(u-\sqrt{\beta }x)H_{n}(u)du.  \nonumber
\end{align}
Eq. \eqref{5.27} can also be written as
\begin{align}<\xi \mid \left( x,y\right) ,\beta ,m>=\sum_{n=0}^{+\infty }&\phi
_{n}\left( \xi \right) \int\limits_{\mathbf{-\infty }}^{+\infty }du\left(
\sqrt{\pi }2^{n}n!\right) ^{-\frac{1}{2}}e^{-\frac{1}{2}u^{2}}H_{n}(u)
\label{5.28} \\
& \qquad \quad\times \left( \left( \sqrt{\pi }2^{m}m!\right) ^{-\frac{1}{2}}e^{i\sqrt{%
\beta }uy-i\frac{\beta }{2}xy-\frac{1}{2}\left( u-\sqrt{\beta }x\right)
^{2}}H_{m}(u-\sqrt{\beta }x)\right) .  \nonumber
\end{align}
From Eq. \eqref{5.28}, we see that the wave function $<\xi \mid
\left( x,y\right) ,\beta ,m>$ defined in \eqref{5.11} has the form:
\begin{equation}
<\xi \mathsf{I}\left( x,y\right) ,\beta ,m>=\left( \sqrt{\pi }2^{m}m!\right)
^{-\frac{1}{2}}e^{-i\sqrt{\beta }\xi y+i\frac{\beta }{2}xy-\frac{1}{2}\left(
\xi -\sqrt{\beta }x\right) ^{2}}H_{m}(\xi -\sqrt{\beta }x).  \label{5.29}
\end{equation}
This ends the proof. \end{proof}

Finally, by the help of Proposition {5.1}, one can use a
little calculation to arrive at the announced formula:

\noindent \textbf{Theorem 5.1. }\textit{Let }$m=0,1,2, \cdots ,\beta >0$\textit{\ and }$%
a,b\in \Bbb{R}$. \textit{Then},
\begin{align}
\sum_{k=-m}^{+\infty }\frac{\left( \sqrt{\beta }\left( a+ib\right) \right)
^{k}}{2^{k}\left( k+m\right) !} & L_{m}^{(k)}\left( \frac{\beta }{2}\left(
a^{2}+b^{2}\right) \right) H_{k+m}\left( \xi \right)  \label{5.30} \\
=&\frac{1}{m!}\exp \left( -\frac{\beta }{4}\left( a-ib\right) ^{2}+\xi
\left( a-ib\right) \sqrt{\beta }\right) H_{m}\left( \xi -a\sqrt{\beta }%
\right)  \nonumber
\end{align}

\noindent \textbf{Remark 5.2.} As mentioned in Section 1, if we take $m=0$, $b=0$ and $
\tau =\frac{\sqrt{\beta }}{2}a,$ then Eq. \eqref{5.30} reduces to
the generating function in \eqref{1.4}.\\

\noindent \textbf{Remark 5.3.} According to Definition 5.2, the coherent states
transform associated with the constructed coherent states have
been used in \cite{17}  to characterize bound states of the planar
Landau problem as integral transforms of square integrable functions on the
real line.

\end{document}